# Ultrashort Ne$^{q+}$ Ion Pulses for Use in Pump-Probe Experiments: Numerical Simulations

P. Kucharczyk [1], A. Golombek [1], M. Schleberger [1], A. Wucher [1] and L. Breuer [1]

[1] Fakultät für Physik, Universität Duisburg-Essen, D-47057, Duisburg, Germany

E-mail: lars.breuer@uni-due.de

**Abstract**

A time resolved experiment to investigate the ultrafast dynamics following an ion impact onto a solid surface requires an ultrashort ion pump pulse in combination with a properly synchronized and time resolved probe. In order to realize such an experiment, we have investigated a strategy to use femtosecond laser photoionization of atoms entrained in a pulsed supersonic jet for the production of sufficiently short ion pulses. While the generation of Ar$^{q+}$ ions was targeted in previous work, it has in the meantime been demonstrated that argon is not suitable due to extensive cluster formation in the supersonic expansion. Here, we therefore present numerical simulations investigating the use of neon as a precursor gas and show the feasibility of pulses containing up to ~1000 Ne$^{q+}$ ions at keV energies and picosecond duration. In the process, we demonstrate that space charge broadening can be significantly reduced by detuning the flight time focusing conditions of an ion bunching system. Moreover, the results show that a controlled variation of the buncher geometry and potentials permits the generation of picosecond pulses at variable ion energy between 1 and 5 keV.

## 1. Inroduction

To follow ultrafast dynamics in solids, i.e. excitation by laser radiation, pump probe schemes have been used with great success [1-6]. These schemes are widely used to directly follow the relaxation of excitation events in the time domain with a temporal resolution of (sub-)picoseconds. To date, there are no comparable experiments to follow the dynamics triggered by the impact of an energetic particle onto a solid surface. The time scales involved in this case are very similar to those for laser excitation. Such a particle, typically an ion, transfers its kinetic energy rapidly (on as – fs time scales) and very locally to the nuclei and electrons of the target, creating an extreme non-equilibrium state. All available information about the equilibration of these states rely on computer simulations and theoretical approaches, which currently can only be experimentally verified by measuring the asymptotic final state. These leads to assumptions and simplifications justified only by the correct description of observables characterizing the state of the experiment rather long times after the impact, such as sputter yields, electron yields, or angular and energy distributions of emitted particles (atoms, electrons or photons). The temporal dynamics of the underlying atomistic processes has not yet been directly accessible by experimental measures. One can imagine probing these underlying processes directly by using the same pump-probe approach as used for investigating other ultrafast dynamics in solids. The challenge here is to produce an ultra-short ion pulse as a pump that must be jitter-free synchronized with a probe pulse, i.e., a femtosecond laser pulse and a well-known impact time onto the target surface.

Several groups have developed different strategies to produce such ion pulses [7-9], involving acceleration in laser driven plasmas or so-called neutralized drift compression techniques [10]. These approaches led to light ions (H and He) with kinetic energies of 1 to 10 MeV [11]. When these ions collide with a solid surface at such energies, the energy transfer process is completely dominated by electronic stopping, leading to an energy transfer from the ion to the electronic system of the target comparable to laser excitation. Since most applications of ion beams operate with heavy ions in the range of nuclear stopping, where the initial pulse leads to a collisional cascade of mostly elastic collisions, this energy range is of great interest. Therefore, ultrashort ion pulses with keV energies are needed. While bunching techniques generally allow to produce short ion pulses down to several ten picoseconds in that regime, these pulses yield broad energy spectra due to the bunching process. Therefore, we have proposed an approach using femtosecond strong field-ionization (SFI) of ultracold atoms from a supersonic neutral gas jet [12]. Earlier experiments utilizing argon at room temperature as a process gas have demonstrated pulse widths on the order 180 ps for keV $Ar^{q+}$ ions. The main limiting factor for the pulse width in this case was thermal broadening due to the thermal motion of the argon atoms prior to the ionization event. While cooling of the argon gas is in principle possible, it yields complications in our experimental approach. Adiabatic expansion of argon into vacuum leads to extensive cluster formation and is used in many commercial ion sources to produce gas cluster ion beams. While these applications rely on the formation of such gas clusters, cluster formation is unwanted in our experimental approach since the energy release after the ionization event counteracts the supersonic cooling effect. The experiment has therefore now been switched to neon as a precursor gas, and we have demonstrated the generation of $Ne^+$ ion pulses from a supersonic beam which are significantly shortened as compared to those produced by room temperature background gas [13]. However, the experimentally measured pulse width of about 135 ps is now limited by the time resolution of the currently used ion detection system, thereby generating the need to perform numerical simulations for this system in order to investigate the prospects of the method.

Due to the greater ionization potential of neon compared to argon, we expect different spatial ionization profiles as well as different space charge interactions as compared to our previous simulations. Additionally, both atom species differ in the rebound momentum induced by the liberated photoelectrons, resulting in different starting velocity distributions of the ions immediately after photoionization. For that reason, it is not possible to directly rescale the results of numerical simulations performed previously for argon ions just in terms of the different ion mass [14]. In order to correctly predict the arrival time distribution for neon ions, one has to specifically model the photoionization process in the same way as done in [13, 14] for argon. For situations where more than one ion is produced per laser shot, the spatial distribution of the resulting ion clouds must be included in the following ion trajectory simulations in order to account for the coulomb interaction within the generated ion bunch. In addition to the strategy to mitigate space charge induced broadening in such situations by blowing up the effective ionization volume [14], we present a technique to further reduce this effect by a controlled detuning of the flight time focusing conditions in the ion bunching system, thereby allowing the generation of picosecond pulses containing more than 1000 $Ne^+$ ions at 5 keV energy. Moreover, we will show that a controlled in-situ variation of the buncher geometry along with its electrode potentials allows to tune the ion energy down to about 1 keV without sacrificing the time resolution.

## 2. Simulations

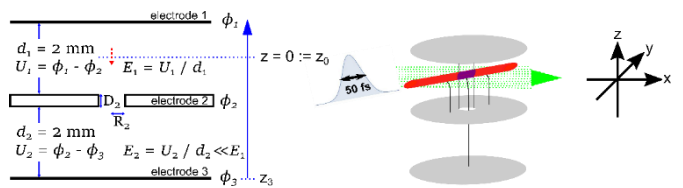

*Figure 1: Left: simulation volume of the ion optical buncher consisting of two (nearly) homogeneous electric fields $E_1$, $E_2$ between electrodes 1 & 2 and 2 & 3, respectively. Middle: sketch of the total ionization volume (bright shaded, red) and the volume from which the ions contributing to the pulse reaching the target electrode 3 must originate (dark shaded, purple), green dotted lines indicate the Ne atoms entrained in the supersonic jet. Right: The coordinate system used in this paper*

Figure 1 (left panel) shows the ion bunching electrode arrangement consisting of three parallel electrodes, where the ions are generated via photoionization using an intense, ultrashort laser pulse, which is tightly focused into the gap between the electrodes 1 and 2 to reach peak power densities up to $10^{17}$ W/cm$^2$ in the focal waist. The generated photo-ions are accelerated through an 80 µm diameter aperture in electrode 2 towards electrode 3 representing the target surface. The center electrode 2 is assumed to have a thickness of $D_2$ = 0.1 mm. The laser peak intensity was always chosen such that the desired number of ions per pulse is generated in the resulting effective ionization volume limited by the diameter of the aperture in electrode 2 as depicted in purple color in Figure 1 (middle panel), and ions generated outside this volume are neglected since they do not contribute to the ion pulse hitting the target electrode 3.

Details of the ion trajectory simulations have been provided in [13, 14] and will not be repeated here. Briefly, the electric field was calculated using the boundary element method implemented in the software package "Charged Particle Optics (CPO)" [15-17], with the initial result being refined in the central region around the ion-optical axis using a finite difference approach similar to the one implemented in the SIMION [18] code. Ion trajectories were followed using the "General Particle Tracer (GPT)" [19-22] software within a static electric field, i.e., the influence of the space charge generated by the flying ion bunch on the potential field was neglected. The starting positions and charge states of the ions were statistically selected according to probability distributions derived from standard Ammosov, Delone, Krainov (ADK) strong-field photoionization theory [23]. Starting velocities of the resulting photo-ions were selected according to a thermal distribution at a given temperature $T_{Start} = (T_\parallel, T_\perp)$, where $T_\parallel$, $T_\perp$ describe the velocity distributions parallel and perpendicular to the propagation direction of the gas beam, respectively. From experimental data [23] we set $T_\parallel$ = 500 mK and $T_\perp$ = 280 mK. The drift velocity due to the gas expansion $v_\parallel$ was assumed as $v_\parallel$ = 770 m/s, again derived from corresponding experimental data [24].

The trajectories of all ions within a bunch were run in parallel with full account of the coulombic interaction between the ions. Only ions transmitted through the aperture in electrode 2 were counted, and the flight time of each ion reaching the target electrode 3 was calculated and histogrammed in bins of usually 20 fs width. In order to assess the temporal pulse width, the resulting histograms are plotted vs. the deviation from the most probable ion arrival time ("relative flight time").

## 3. Results and Discussion

The coordinate system used in the remainder of this paper is shown in Figure 1 (right panel). The supersonic beam and laser beam propagate in *x*- and *y*- direction, respectively, and the photo-ions are accelerated in (negative) *z*-direction.

### 3.1 Optimizing electric fields

In ref. [14] we have shown that space charge broadening can be significantly reduced by using an astigmatic laser beam profile for photoionization of the atoms entrained in the gas jet. In this chapter we present a technique to further reduce space charge broadening by a controlled detuning of the flight time focusing conditions. Assuming the potentials $\phi_1$ = 0 V and $\phi_3$ = -10 kV for the electrodes 1 and 3, the potential $\phi_2$ at the center electrode 2 is normally fixed by the requirement to ensure first order flight time focusing conditions for ions generated in the center between electrodes 1 and 2 at the position of the target surface 3. For ion pulses containing only one single ion, this configuration results in the temporally narrowest ion arrival time distribution. In this case, the ion "pulse width" is determined by the FWHM of this distribution. The picture changes if one increases the number of ions per pulse, for instance by increasing the laser intensity. In this case, the space charge interaction between the ions results in a significant broadening of the ion arrival time distribution. The temporal width (FWHM) of ion pulses containing a variable number of ions is shown as a function of the potential $\phi_2$ of the center electrode Figure 2. Under "normal" conditions, this potential would be kept fixed at the value of $\phi_2$ = -9652.5 V, which ensures first order flight time focusing conditions for single ions starting around the center of the effective ionization volume (indicated by the dotted line). It is evident that this would lead to an extensive broadening of the pulse if the number of ions is increased.

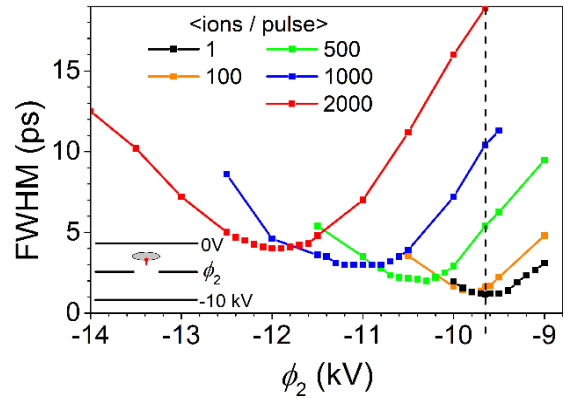

*Figure 2: FWHM of flight time distributions calculated for Ne+ ion pulses containing a variable number of ions per pulse. The data were generated assuming an astigmatic laser beam ($w_{0x}$ = 72 μm, $w_{0z}$ = 15.25 μm) and a gas beam density of $10^{17}$ m$^{-3}$. The laser intensity $I_0$ is being adjusted to deliver the desired number of ions.*

On the other hand, a controlled detuning of the flight time focusing conditions via variation of $\phi_2$ can obviously reduce this effect. The optimum value of $\phi_2$, leading to the temporally shortest ion pulses, depends on the number of ions per pulse and shifts with increasing number of ions. As seen from Figure 2, the duration of pulses containing 2000 ions can be reduced by about 75 % down to about 4 ps using this strategy. The reason for the apparent re-compression of the pulse by deviating from first order flight time focusing conditions becomes more clear in Figure 3, which shows the flight time distributions of the ions calculated for this example in the top row along with the corresponding distribution of their starting positions in the plane perpendicular to the laser propagation axis (bottom row). In order to visualize the space charge influence, each starting position is colored by the respective relative ion flight time for different potentials $\phi_2$. Under first order flight time conditions ($\phi_2$ = -9652.5 V), single ions which start at the center of the effective ionization volume would exhibit the shortest flight time, while ions starting on either side away from this point would exhibit the same flight time deviation ("flight time focusing"). If the ion packet contains more than one ion, on the other hand, ions starting on the side closer to the target electrode 3 are additionally accelerated along the extraction direction by the space charge of the ion cloud, thereby leading to shorter flight times, while ions generated further away from electrode 3 are accelerated in the opposite direction, thus leading to longer flight times. This correlation between ion velocity and starting position within the travelling ion packet can be utilized in order to at least partly mitigate the influence of space charge broadening via a controlled detuning of the flight time focusing properties. More specifically, the variation of $\phi_2$ generates a flight time dispersion as a function of starting position along the extraction axis which is opposed to the space charge induced dispersion. This way, it is possible to realize pulses containing more than $10^3$ ions at pulse durations of several picoseconds as shown in Figure 2. A further variation of $\phi_2$ finally overcompensates the space charge effects, resulting in the fact that the ions generated closer to electrode 3 become the slowest ions in the bunch. The corresponding flight time distribution of the ion pulses is shown in the top row of Figure 3, illustrating the effect of counteracting space charge induced pulse broadening by appropriate variation of $\phi_2$.

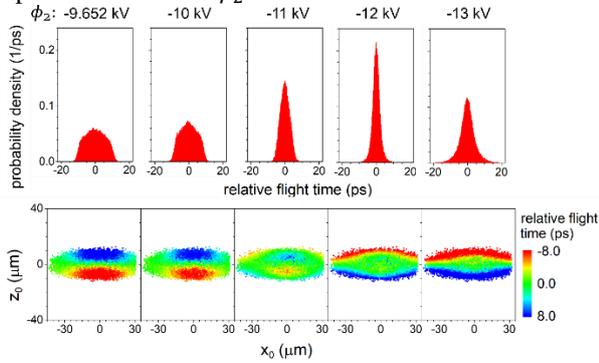

Figure 3: top) Relative flight time distributions calculated for Ne$^+$ ions within a pulse containing on average 2000 ions at different values of the potential $\phi_2$. Bottom) Flight time of ions reaching the target electrode 3 vs. starting position in the xz-plane located 3.1 mm above electrode 3 corresponding to the flight time distributions depicted in the top row.

*3.2 Optimizing electrode distances*

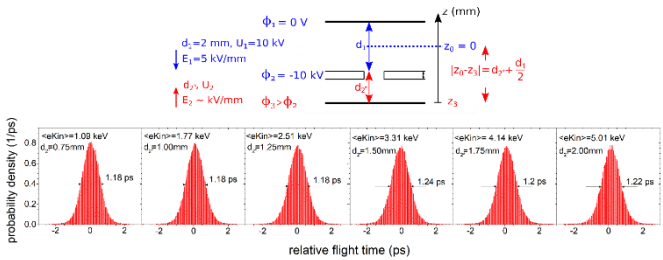

Figure 4: top) sketch of the simulated buncher geometry. Quantities during variation of $d_{2*}$ are depicted either blue (constant, while $d_{2*}$ is varied) or red (varied with $d_{2*}$). bottom) relative flight time distributions for Ne$^+$ pulses containing 10 ions on average with different kinetic energies per ion. The parameter in the calculation were $I_0 = 5.1 \times 10^{14}$ W/cm², a laser focus diameter of 10 µm and a neutral gas density of $10^{17}$ atoms/m³.

The temporal broadening of ion pulses due to their initial velocity distribution (parametrized by the temperature $T_\perp$) along the extraction field $\vec{E}_1$ can be estimated from the "turn-around-time" difference of ions starting in opposite directions along the extraction axis as [25]:

$$FWHM_{ions}(|\vec{E}_1|) = \frac{2m}{q \cdot |\vec{E}_1|} \cdot \sqrt{2 \cdot ln(2)} \cdot \sqrt{\frac{k_B \cdot T_\perp}{m}} \propto \frac{1}{|\vec{E}_1|}$$

According to this estimate the pulse length is determined by the electric field strength $\vec{E}_1$ between the top electrodes. Reducing the kinetic energy of the ions at the target electrode 3 by reducing the electric field strength $\vec{E}_1$ therefore leads to a broadening of the ion pulse $\propto 1/|\vec{E}_1|$. On the other hand, it would be desirable to reduce the kinetic impact energy of the ions without broadening the temporal pulse length. A possibility to generate such ion pulses is by adjusting the buncher as depicted in Figure 4 top). In this configuration, the potentials $\phi_1 = 0$ V, $\phi_2 = -10$ kV and the distance $d_1 = 2$ mm are fixed, resulting in a fixed extraction field $|\vec{E}_1| = 5$ kV/mm. As the ions are generated in the center of the electrodes 1 and 2, while the distance $d_{2*} = D_2 + d_2$ is reduced, flight time focusing is possible if an appropriate electric field $\vec{E}_2$ is applied, which is opposed to $\vec{E}_1$. This way the electric field $|\vec{E}_1| = 5$ kV/mm leads to a strong compression of the ions between the electrodes 2 and 3, while $\vec{E}_2$ simultaneous deaccelerates the ions. This results in ion pulses with reduced kinetic energy at the target electrode 3. By varying the distance $d_2$ in the range 0.75 mm – 2.0 mm, and adjusting the electric field $\vec{E}_2$ in order to ensure first order flight time focusing conditions at the target surface, the resulting arrival time distribution of ions in pulses containing on average 10 ions are shown in Figure 4 (bottom). It is seen that the pulse length attainable by this strategy is nearly constant around 1.2 ps and determined almost exclusively by the temperature $T_\perp$ parametrizing the ion velocity distribution along the extraction axis. A correlated change of the target electrode potential along with a corresponding in-situ change of its position via, for instance, a motorized micrometer stage, would therefore open the possibility to continuously vary the ion energy in the range between 1 and 5 keV without any sacrifice to the picosecond pulse duration. A corresponding requirement for the experimental setup, namely the possibility to move electrode 3 while the experiment is running, is therefore incorporated in our new setup.

## 4. Conclusions

We present detailed ion trajectory simulations to calculate the arrival time distributions of Ne$^+$ ions generated via photoionization in a cold supersonic neon beam with full account of their generation, the resulting start configuration with respect to charge state, position and velocity distributions as well as the space charge broadening. In particular, we show that the generation of pulses containing from single up to several thousand ions with energies in the keV regime and picosecond duration is possible. By a controlled detuning of the flight time focusing conditions of the ion optical bunching setup, we found a useful method to counteract the effect of pulse broadening due to space charge effects. Moreover the simulations reveal that it is possible to freely adjust the kinetic ion energy by appropriate adjustments of the buncher potentials and geometry.

In order to verify these predictions experimentally, it is necessary to improve the time resolution of our ion detection system from currently 135 ps [13]. For that purpose, we are working to replace the currently used microchannel plate (MCP) ion detector by a streak technique delivering sub-picosecond resolution. Ultimately, it is planned to use the generated ion pulses in order to study the nuclear and electronic dynamics following an ion impact onto the target surface and compare the results with theoretical model calculations based, for instance, on molecular dynamics simulations. Possible probe strategies in such an ion pump - laser probe experiment include time resolved X-ray or electron diffraction, photoelectron emission or photoionization to study ion-induced order-disorder transitions, transient electronic excitations or nuclear collision and particle emission dynamics. The advantage of the laser based ion pulse generation is that the laser-based probe can be optically synchronized with the ion impact without any electronic triggering jitter. We are confident that the strategy described here will therefore open path for a long missing experimental access to the ion-induced dynamics in solids.

## Acknowledgements

This work was funded by the Deutsche Forschungsgemeinschaft (DFG, German Research Foundation - Projektnummer 278162697 - SFB 1242) in the frame of the project "Particle-Induced Excitations" (C05) within the Collaborative Research Center (SFB) 1242 "Non-equilibrium dynamics in the time domain".